\documentclass{article}

\usepackage[square,sort,comma,numbers]{natbib}

\usepackage[preprint]{neurips_2023}

\usepackage[utf8]{inputenc} 
\usepackage[T1]{fontenc}    
\usepackage{hyperref}       
\usepackage{url}            
\usepackage{booktabs}       
\usepackage{amsfonts}       
\usepackage{nicefrac}       
\usepackage{microtype}      
\usepackage{xcolor}         
\usepackage{multirow}
\usepackage{graphicx} 
\usepackage{amsmath} 

\title{Efficient Data Distribution Estimation for Accelerated Federated Learning}

\newcommand{\iid}{IID }
\newcommand{\py}{$P(y)$ }

\newcommand{\pxy}{$P(X \mid y)$ }
\newcommand{\pxyy}{$P(X \mid y)$}

\author{%
  Yuanli Wang \\
  Boston University\\
  \texttt{yuanliw@bu.edu} \\
  \And
  Lei Huang \\
  Boston University\\
  \texttt{lei@bu.edu} \\
}

\begin{document}

\maketitle



\section{Introduction}

Federated Learning(FL) is a privacy-preserving machine learning paradigm where a global model is trained in-situ across a large number of distributed edge devices. These systems are often comprised of millions of user devices and only a subset of available devices can be used for training in each epoch. Designing a device selection strategy is challenging, given that devices are highly heterogeneous in both their system resources and training data. This heterogeneity makes device selection very crucial for timely model convergence and sufficient model accuracy \cite{edgesys21}. 
To tackle the FL client heterogeneity problem, various client selection algorithms have been developed, showing promising performance improvement in terms of model coverage and accuracy \cite{fu2023client, acmeflsurvey}. 

In this work, we study the overhead of client selection algorithms in a large scale FL environment. Then we propose an efficient data distribution summary calculation algorithm to reduce the overhead in a real-world large scale FL environment.
The evaluation shows that our proposed solution could achieve up to 30x reduction in data summary time, and up to 360x reduction in clustering time.


\section{Heterogeneity-aware cluster based Federated Learning}

\begin{figure}[t]
\centering
\includegraphics[width = 0.96\textwidth]{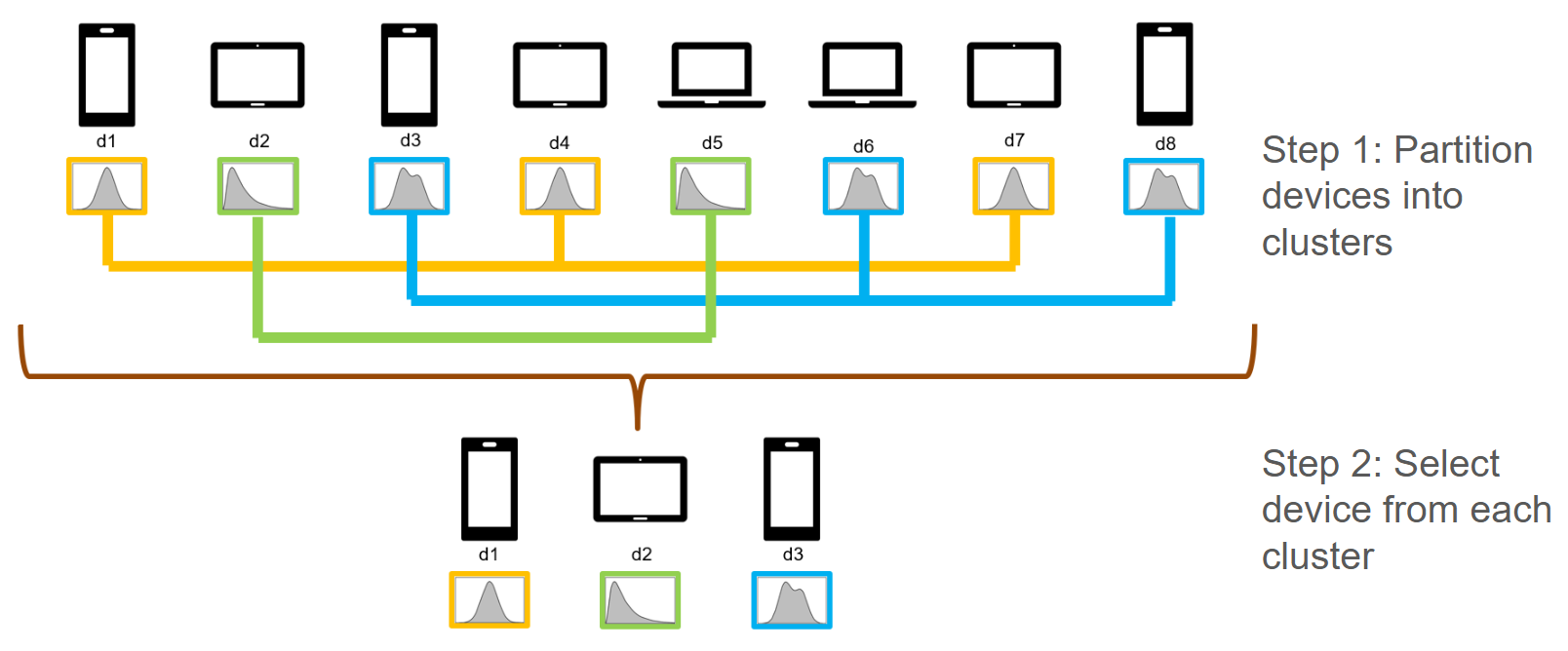}
\caption[center]{system overview and workflow of query deployment.} 
\label{sys_arch}
\vspace{-0.3cm}
\end{figure}

In FL, data is not independent and identically distributed (IID) across all edge devices, resulting in statistical heterogeneity across devices \cite{wang2020poster}. Many existing works \cite{haccs, ai3010008, pmlr-v139-fraboni21a, kdd20fls, acmeflsurvey, pan2023contextual} exploit client’s local data to quantify the statistical heterogeneity, and cluster client devices based on their discernible statistical differences. Then for every training iteration, they select a cluster of devices based on client's system heterogeneity and statistical heterogeneity. As an example, Figure~\ref{sys_arch} shows the workflow of HACCS\cite{haccs}.

HACCS\cite{haccs} proposed two types of data-sampling based \textbf{\emph{distribution summaries}} to quantify the statistical heterogeneity, motivated by techniques for identifying \iid violations. Standard distributed machine learning models assume that the response variables $y_i$ and the predictor variables $X_i$ at each device are drawn \iid from a shared joint distribution, $P(X, y)$. This joint distribution can be factored as follows:
\begin{equation} \label{eq:joint}
    P(X, y) = P(y) \; P(X \; | \; y)
\end{equation}
Therefore, if $P(y)$ (the marginal distribution of the response labels) or $P(X \; | \; y)$ (the data distribution conditioned on the response) differs at any device, we have a violation of the \iid assumption. When \iid violations occur, it follows that \textit{one of these distributions must differ across one or more devices} \cite{lit-book}. Based on this insight, they propose $P(y)$ and $P(X \; | \; y)$ as distribution summaries. 
Their evaluation on a 50 clients setting with FEMNIST and CIFAR-10 datasets shows that \py and \pxy can provide 18\%-38\% reduction in  model training time compared to the state of the art, without any compromise in accuracy.

\subsection{Limitations}

HACCS\cite{haccs} only computes distribution summary of each client once in the first training iteration. Thus, the overhead of computing distribution summary has negligible impact on overall training speed. 
However, in the real case, as users' application running, the data distributions of the clients may be time-varying and non-stationary \cite{ganguly2023online, zhang2023federated}. In order to do adaptive client selection, we need to re-compute distribution summary periodically as data changes.
For slow edge computing devices with a larger scale of data samples, computing distribution summaries is time-consuming with high memory consumption.
Also, when the number of clients becomes large, clustering a large amount of client distribution summaries will take a long time.
The overhead of these algorithms may become large and slow down training. 

In FL setting, devices also have system heterogeneity, which means devices have different processing capacity, network bandwidth, and power. Since the available resources of each device change rapidly, we need each device to periodically send its resource status to the central server. We propose efficient data distribution summary algorithm that enables devices to periodically update their distribution summary, which could enable more robust FL system in real world environment.


\section{Motivation study}

\subsection{Dataset and motivation study}

To motivate our solution, we first perform an empirical evaluation study to quantify the overhead of computing distribution summaries in existing work ~\cite{haccs}. The results of our study motivate the design of our proposed efficient data distribution summary algorithm. 

We use two large scale real-world datasets in Table~\ref{table-dataset} to evaluate \pxyy and \py distribution summary methods. The datasets are pre-processed and partitioned to different clients by \cite{lai2022fedscale}.

\begin{table}
\centering
\vspace{2mm}
\begin{tabular}{|p{1.5cm}|p{2.4cm}|p{1.9cm}|p{1.4cm}|p{1.4cm}|p{2.1cm}|}
\hline
Dataset & Application & sample size & number of classes & number of clients & number of samples per client    \\ \hline
FEMNIST & handwritten alphanumeric classification & 28x28x1 grayscale images & 62 & 2800 & average: 109 \newline max: 6709 \newline std: 211.63 
   \\ \hline
OpenImage & image classification & 3x256x256 color images & 600 & 11325 & average: 228 \newline max: 465 \newline std: 89.05    \\ \hline
\end{tabular}
\vspace{2mm}
\caption{Datasets used for evaluation}
\label{table-dataset}
\end{table}

Table ~\ref{tabresults} shows the overhead of two existing distribution summary methods \pxy and \py. We found that for larger dataset, \pxyy  needs up to 553s to finish calculating summary, which is quite slow. Also, we monitored that it uses more than 64GB memory for computation, which is not acceptable for mobile devices that typically have less than 16 GB memory. 
Although \py can efficiently compute on large dataset, it fails to capture the data heterogeneity of the feature of samples under the same label, since it only considers the label distribution  
(For example, images of both cats and dogs might be labeled as "animals", but their features could be quite different).

We also find that since the size of summary becomes large, it will take long time to do clustering, even up to more than 2 days. Furthermore, the cluster algorithm (DBSCAN) is sensitive to parameter setting. When we reuse the parameters tuned for one dataset to another setting, it can sometimes put all devices to the same group, and can not return a meaningful clustering solution.


\section{Proposed Solution}

\subsection{Distribution summary calculation}

We propose to use dimension reduction and coreset to allow efficient calculation of the distribution summary on each device. In this way we can both accelerate the computation time and reduce the size of the summary. 
For each device, we construct the coreset by sampling $k$ elements from the dataset on this device, while maintaining its original label proportions. 
Then for each element in the coreset, we use encoder to do dimension reduction on its feature. Specifically, we modified MobileNet\cite{mobilenetv3} network and extract the output of a hidden layer as the feature vector. The advantage of using encoder rather than PCA or Johnson–Lindenstrauss Lemma is: (1). the encoding computation could be easily accelerated by GPU. (2). MobileNet is pre-trained for image classification tasks and can capture the spatial information of pixels on the image (feature).

After coreset selection and dimension reduction, for each device, we construct the distribution summary as a flat vector that includes: (1). the element-wise mean of feature vectors of samples under each unique label.   (2). the distribution of each unique label. As the result, suppose the output of dimension reduction is a vector with size $H$, the number of classes is $C$, then the shape of distribution summary will be $C*H+C$. It is much smaller than the histogram representation used in \pxyy, but still included the distribution information of both sample features and labels.

\subsection{Device clustering}

We use K-means for device clustering, since it fits to our simplified distribution summary. K-means is a popular clustering algorithm that partitions a dataset into $k$ distinct, non-overlapping clusters. The target of the K-means clustering is to minimize the variance within each cluster as denoted by:
\begin{equation}
\text{minimize} \quad J = \sum_{j=1}^{k} \sum_{i=1}^{n} \left\| x_i^{(j)} - c_j \right\|^2
\end{equation}

The objective function $J$ is the sum of distances between all data points to their respective centroids. The procedure of K-means is as follows: (i) it first randomly defines $k$ centroids and assigns each data point to its closest centroid forming $k$ clusters. (ii) For each cluster, it obtains the new centroid by calculating the mean position among all data points within the cluster. It then repeats the step (i) to form new $k$ clusters. (iii) Repeat the previous 2 steps until convergence or a termination condition is reached (e.g. centroids do not change anymore).

K-means is an effective clustering method that minimizes the variance within each cluster and guarantees explicit $k$ partitions of the data set.


\section{Preliminary Results and Future works}

We implemented our solution into the FL framework proposed in HACCS \cite{haccs}. Our proposed solution is complementary to privacy-preserving methods that could be applied on the data summaries, such as differential privacy used in HACCS \cite{haccs}. 
Table~\ref{tabresults} compares the speed of computing data distribution summary on the two datasets. 
We can see that our proposed solution could achieve up to 30x reduction in data summary time, and up to 360x reduction in clustering time.

\begin{table}
\centering
\begin{tabular}{|l|llll|ll|}
\hline
    & \multicolumn{4}{l|}{Time(s) calculating summary}                                                                                     & \multicolumn{2}{l|}{Time(s) clustering devices}                            \\ \hline
    & \multicolumn{2}{l|}{OpenImage}                                              & \multicolumn{2}{l|}{FEMNIST}                           & \multicolumn{1}{l|}{\multirow{2}{*}{OpenImage}} & \multirow{2}{*}{FEMNIST} \\ \cline{1-5}
    & \multicolumn{1}{l|}{Avg}             & \multicolumn{1}{l|}{Max}             & \multicolumn{1}{l|}{Avg}             & Max             & \multicolumn{1}{l|}{}                           &                          \\ \hline
\py    & \multicolumn{1}{l|}{\textless{}0.01} & \multicolumn{1}{l|}{\textless{}0.01} & \multicolumn{1}{l|}{\textless{}0.01} & \textless{}0.01 & \multicolumn{1}{l|}{\textcolor{red}{835.69}}                     & 24.5                     \\ \hline
\pxy   & \multicolumn{1}{l|}{9.13}            & \multicolumn{1}{l|}{\textcolor{red}{553.41}}          & \multicolumn{1}{l|}{3.23}            & 6.68            & \multicolumn{1}{l|}{\textcolor{red}{more than 2 days}}                & \textcolor{red}{1866.40}                  \\ \hline
\textcolor{teal}{Encoder+Kmeans}   & \multicolumn{1}{l|}{\textcolor{teal}{4.6}}             & \multicolumn{1}{l|}{\textcolor{teal}{18.7}}            & \multicolumn{1}{l|}{\textcolor{teal}{5.4}}             & \textcolor{teal}{11.2}            & \multicolumn{1}{l|}{\textcolor{teal}{477.2}}                      & \textcolor{teal}{30}                       \\ \hline
\end{tabular}
\vspace{2mm}
\caption{Overhead comparison of different summary algorithms}
\label{tabresults}
\end{table}

In a large-scale FL setting, the size of the data summary impacts both the network bandwidth usage between clients and the server, and the memory consumption on the server during clustering tasks.
For future work, we plan to explore additional dimension reduction methods to more effectively compress the data summary while maintaining the integrity of statistical diversity information.

\bibliographystyle{plain}
\bibliography{reference}


\end{document}